\documentclass[%
reprint, amsmath,amssymb, showpacs,superscriptaddress, aps,prb,
]{revtex4-2}
\usepackage{graphicx}
\usepackage{dcolumn}
\usepackage{bm}
\usepackage[mathlines]{lineno}
\usepackage[usenames,dvipsnames]{color}
\usepackage{subfig}
\usepackage{color}
\usepackage{float}
\usepackage{subfloat}
\usepackage{caption}
\usepackage[]{graphicx}
\usepackage{dcolumn}
\usepackage{bm}
\usepackage{multirow}
\usepackage{url}
\usepackage{gensymb}
\usepackage{setspace}
\usepackage{dsfont}

\usepackage{amsfonts}
\usepackage{booktabs}
\usepackage{siunitx}
\usepackage{dcolumn}

\newcommand{\cmmnt}[1]{\ignorespaces}

\begin{document}

\title{All-optical spin injection in silicon revealed by element specific time-resolved Kerr effect}

\author{Simone Laterza}
\affiliation{Elettra Sincrotrone Trieste S.C.p.A. Strada Statale 14 - km 163.5 in AREA Science Park 34149 Basovizza, Trieste, Italy}
\affiliation{Department of Physics, University of Trieste, Via A. Valerio 2, 34127 Trieste, Italy}
\author{Antonio Caretta}
\email{antonio.caretta@elettra.eu}
\affiliation{Elettra Sincrotrone Trieste S.C.p.A. Strada Statale 14 - km 163.5 in AREA Science Park 34149 Basovizza, Trieste, Italy}
\author{Richa Bhardwaj}
\affiliation{Elettra Sincrotrone Trieste S.C.p.A. Strada Statale 14 - km 163.5 in AREA Science Park 34149 Basovizza, Trieste, Italy}

\author{Roberto Flammini}
\affiliation{ISM-CNR, Via del Fosso del Cavaliere 100, 00133, Roma, Italy}
\author{Paolo Moras}
\affiliation{Istituto di Struttura della Materia-CNR (ISM-CNR), Strada Statale 14 - km 163.5 in AREA Science Park 34149 Basovizza, Trieste, Italy}
\author{Matteo Jugovac}
\affiliation{Istituto di Struttura della Materia-CNR (ISM-CNR), Strada Statale 14 - km 163.5 in AREA Science Park 34149 Basovizza, Trieste, Italy}
\author{Piu Rajak}
\affiliation{CNR-IOM, Strada Statale 14 - km 163.5 in AREA Science Park 34149 Basovizza, Trieste, Italy}
\author{Mahabul Islam}
\affiliation{CNR-IOM, Strada Statale 14 - km 163.5 in AREA Science Park 34149 Basovizza, Trieste, Italy}
\author{Regina Ciancio}
\affiliation{CNR-IOM, Strada Statale 14 - km 163.5 in AREA Science Park 34149 Basovizza, Trieste, Italy}

\author{Valentina Bonanni}
\affiliation{Elettra Sincrotrone Trieste S.C.p.A. Strada Statale 14 - km 163.5 in AREA Science Park 34149 Basovizza, Trieste, Italy}
\author{Barbara Casarin}
\affiliation{Department of Physics, University of Trieste, Via A. Valerio 2, 34127 Trieste, Italy}
\author{Alberto Simoncig}
\affiliation{Elettra Sincrotrone Trieste S.C.p.A. Strada Statale 14 - km 163.5 in AREA Science Park 34149 Basovizza, Trieste, Italy}
\author{Marco Zangrando}
\affiliation{Elettra Sincrotrone Trieste S.C.p.A. Strada Statale 14 - km 163.5 in AREA Science Park 34149 Basovizza, Trieste, Italy}
\affiliation{CNR-IOM, Strada Statale 14 - km 163.5 in AREA Science Park 34149 Basovizza, Trieste, Italy}

\author{Primo{\v z} Rebernik Ribi{\v c}}
\affiliation{Elettra Sincrotrone Trieste S.C.p.A. Strada Statale 14 - km 163.5 in AREA Science Park 34149 Basovizza, Trieste, Italy}
\author{Giuseppe Penco}
\affiliation{Elettra Sincrotrone Trieste S.C.p.A. Strada Statale 14 - km 163.5 in AREA Science Park 34149 Basovizza, Trieste, Italy}
\author{Giovanni De Ninno}
\affiliation{Elettra Sincrotrone Trieste S.C.p.A. Strada Statale 14 - km 163.5 in AREA Science Park 34149 Basovizza, Trieste, Italy}
\author{Luca Giannessi}
\affiliation{Elettra Sincrotrone Trieste S.C.p.A. Strada Statale 14 - km 163.5 in AREA Science Park 34149 Basovizza, Trieste, Italy}

\author{Alexander Demidovich}
\affiliation{Elettra Sincrotrone Trieste S.C.p.A. Strada Statale 14 - km 163.5 in AREA Science Park 34149 Basovizza, Trieste, Italy}
\author{Miltcho Danailov}
\affiliation{Elettra Sincrotrone Trieste S.C.p.A. Strada Statale 14 - km 163.5 in AREA Science Park 34149 Basovizza, Trieste, Italy}

\author{Fulvio Parmigiani}
\affiliation{Elettra Sincrotrone Trieste S.C.p.A. Strada Statale 14 - km 163.5 in AREA Science Park 34149 Basovizza, Trieste, Italy}
\affiliation{International Faculty, University of Cologne, 50937 Cologne, Germany}
\author{Marco Malvestuto}
\affiliation{Elettra Sincrotrone Trieste S.C.p.A. Strada Statale 14 - km 163.5 in AREA Science Park 34149 Basovizza, Trieste, Italy}
\affiliation{CNR-IOM, Strada Statale 14 - km 163.5 in AREA Science Park 34149 Basovizza, Trieste, Italy}

\date{\today}

\begin{abstract}
Understanding how a spin current flows across metal-semiconductor interfaces at pico- and femtosecond timescales has implications for ultrafast spintronics, data processing and storage applications. 
However, the possibility to directly access the propagation of spin currents on such time scales has been hampered by the simultaneous lack of both ultrafast element specific magnetic sensitive probes and tailored metal-semiconductor interfaces. 
Here, by means of free electron laser-based element sensitive Kerr spectroscopy, we report direct experimental evidence of spin currents across a Ni/Si interface in the form of different magnetodynamics at the Ni M$_{2,3}$ and Si L$_{2,3}$ absorption edges. 
This further allows us to calculate the propagation velocity of the spin current in silicon, 
which is on the order of 0.2~nm/fs. 
\end{abstract}

\maketitle

\section{\label{sec:intro}Introduction}
Independent control of charge and spin carriers is the main goal in the field of spintronics. 
Spin-based electronics has been compared favourably to present electronics in terms of switching energy and speed \cite{awschalom07, wang13}; 
spin currents are believed to flow nearly dissipationless \cite{murakami03} and spin systems show coherence times that are larger with respect to charge confinement \cite{huang07}. 
While the flow of charges is readily manipulated by voltages, the control of spin currents still represents a challenge. 
In this context, the injection of the so-called superdiffusive spin currents (SCs) within metals \cite{battiato10} represented an important achievement \cite{eschenlohr13}. 
Yet, in order to integrate spintronic devices with present semiconductor technology, it is required to demonstrate the injection of spins through metal/semiconductor interfaces. 
SCs are generated inside ferromagnetic metals by the use of ultrashort infrared (IR) pulses \cite{battiato16, battiato17}. 
The IR pulse absorption creates a spin-preserving out-of-equilibrium hot electron distribution \cite{battiato10}. 
Since the excited carrier lifetime and velocity are, consistently in all ferromagnetic metals, much larger for spin majority electrons than for spin minority, within few hundreds of femtoseconds a diffusion process of strongly spin-polarized excited carriers is set, leading to an ultrashort spin current pulse. 
Nowadays the injection of SCs from nickel into metals like Au, Fe or Co-Pt multilayer is well established \cite{rudolf12, kampfrath13, seifert16}. 
Conversely, the spin injection into semiconductors presents new challenges, as carriers need to overcome the band gap to enter the semiconductor conduction band. 
Nonetheless, the process is still possible and a recent theoretical work \cite{battiato16} proposed the spin injection from nickel to silicon to be i) chargeless, i.e., independent from the charge carrier flow; ii) up to 80\% spin polarized, and finally iii) ultrashort. 
To the best of our knowledge, there is still no experimental proof of ultrafast spin injection into silicon. 
Silicon, besides its role in semiconductor technology, allows long lived spin currents due to its small spin-orbit interaction, the reduced mean nuclear spin and the crystal inversion symmetry \cite{Tyryshkin03, Zutic04}. 
Also, the inspection of mechanism of SCs at metal/semiconductor interfaces is vital for further advancement in spintronics.\\
\begin{figure*}[ht]
\captionsetup{justification=centerlast}
\centering
\includegraphics[width=0.9\textwidth]{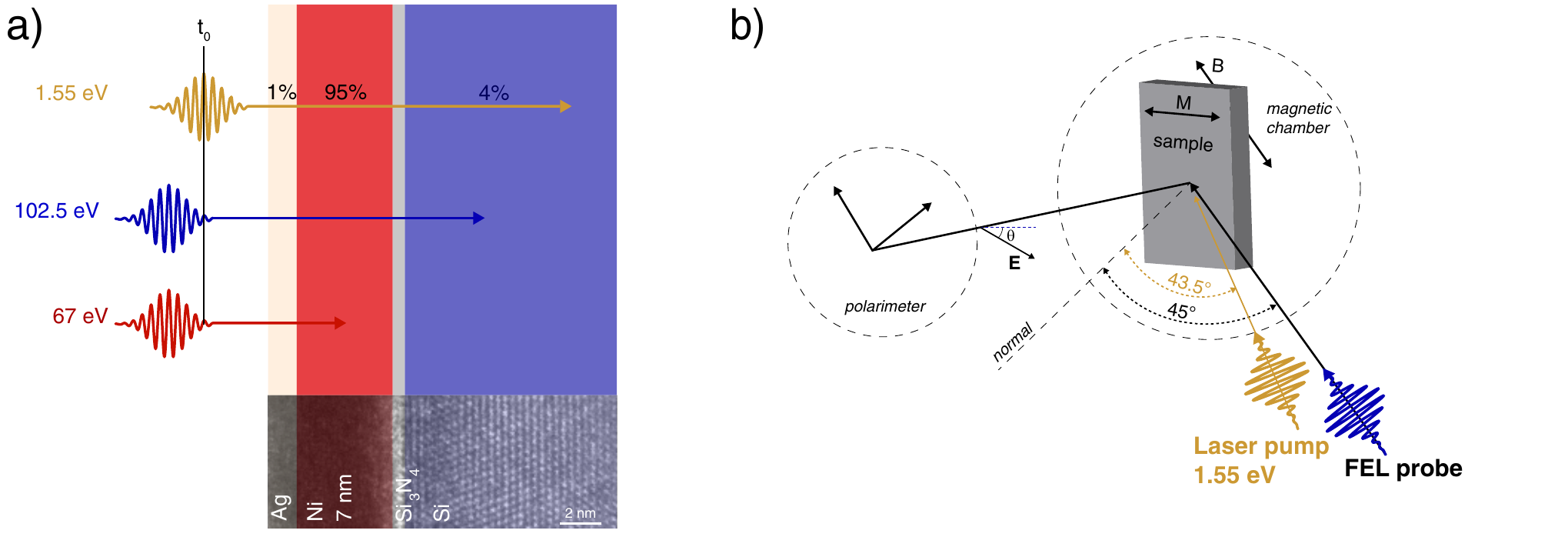}
\caption{\label{fig1}a) Schematic representation of the sample stack consisting of a Si substrate (blue), a thin Si$_3$N$_4$ passivation layer (grey), 7~nm of Ni film (red) and 2~nm of Ag capping layer (yellow). 
At time $t_0$ the IR laser pulses (1.55~eV, gold arrow) hit the sample. 
The FEL pulses, tuned at the Si L$_{2,3}$ edge  (102.5~eV, blue arrow) and at the Ni M$_{2,3}$ edge  (67~eV, red arrow), probe the sample status at later timescales. 
The penetration depth of the pump and of the probe pulses is represented by the length of the respective arrows. 
The fraction of the total energy of the IR laser absorbed by each layer is reported above the pump arrow in percent. 
An HRTEM image of the heterostructure is superimposed to the lower part of the schematic (see section \ref{sec:suppleB}). 
b) Scheme of the FEL RMOKE set-up at MagneDyn. 
The pump and probe pulses are in a quasi collinear configuration.
RMOKE was probed in longitudinal configuration at an angle of incidence of 45$^{\circ}$.
The linear polarization \textbf{E} of the FEL pulses reflected from the sample is rotated form the scattering plane (dashed line) of the Kerr angle $\theta$.}
\end{figure*}
In this work, we provide experimental evidence of the metal-to-semiconductor spin injection in a Ni/Si interface and put forth an insight into the mechanism by comparing ultrafast demagnetization measurements performed using the time resolved resonant magneto-optical Kerr effect (RMOKE) \cite{yamamoto17} at Si L$_{2,3}$ and Ni M$_{2,3}$ absorption edges. 
Thanks to the site selectivity, RMOKE allowed us to decouple the magnetodynamical response of the Si and Ni layers. 
We observed a slower demagnetization response at the Si edge compared to the Ni edge and interpret this difference as an evidence of a transient spin current from the ferromagnetic layer into silicon. 

\section{\label{sec:exp}Experimental}
We investigated a Ni/Si$_3$N$_4$(0001)/Si(111) interface whose structure is schematically represented in Fig.~\ref{fig1}(a). 
The sample was synthesized at the VUV-Photoemission beamline (Elettra Sincrotrone Trieste) according to the recipe of Ref.s~\cite{Ahn01,Flammini15}. 
Section \ref{sec:suppleA} reports on the spectroscopic characterization of the sample. 
The Ni/Si interface has been realized by depositing a 7~nm layer of Ni on top of a p-doped (B dopant, 0.05~$\Omega \cdot$cm resistivity) Si(111)-7x7 surface reconstructed substrate. 
Nitride passivation of the Si surface was required in order to avoid the formation of unwanted metallic silicides and diminish the migration of the metallic ions into the substrate. 
Finally, a Ag capping layer was deposited in order to avoid the oxidation of the Ni layer. 
The Ni/Si$_3$N$_4$/Si(111) nanostructure was characterized by high resolution transmission electron microscopy (HRTEM). 
A representative bright-field HRTEM image of the Ni/Si$_3$N$_4$/Si(111) heterostructure is shown in Fig.~\ref{fig1}(a). 
Further details on HRTEM measurements are provided in section \ref{sec:suppleB}. 
\\
The room-temperature transient magnetic response of our sample was measured via the longitudinal RMOKE in a pump-probe scheme, which allows one to detect the effects on the sample magnetization \textbf{M} induced by an IR pump beam. 
All the RMOKE measurements were carried out at the MagneDyn end-station \cite{Svetina16} at the externally seeded EUV free-electron laser (FEL) FERMI \cite{Allaria.2013} at Elettra Sincrotrone Trieste.
\\
A schematic illustration of the experimental scattering configuration is  shown in Fig.~\ref{fig1}(b). 
The sample was excited by a $\sim$70 fs pump pulse at 1.55~eV with 25~Hz repetition rate decimated with respect to the FEL 50~Hz repetition rate for achieving the standard pump-on/off data acquisition mode. 
The IR pump pulse was generated from the same laser used to seed the FERMI FEL and had a root-mean-square timing jitter with respect to the FEL pulses of $\sim$7 fs \citep{Danailov14}.
The angle of incidence of the incoming IR pump pulse is 43.5\degree. 
The probe consists of $\sim$50 fs FEL light pulses whose energy is tuned to the absorption edges of Ni and Si. 
The electric field of the linearly polarized incoming light lays in the scattering plane, while the angle of incidence was set to 45\degree. 
The spot size and fluence of the IR pump pulses at the sample were $\sim$60$\times$60 \si\micro m$^{2}$ and 120~mJ/cm$^{2}$ for the measurements at the Ni M$_{2,3}$ edge and $\sim$140$\times$110 \si\micro m$^{2}$ and 30~mJ/cm$^{2}$ for the Si L$_{2,3}$ edge. 
Instead, the FEL spot size and fluences were $\sim$54$\times$60 \si\micro m$^{2}$ and 8.0~mJ/cm$^{2}$ for the Ni M$_{2,3}$ edge and $\sim$61$\times$60 \si\micro m$^{2}$ and 2.3~mJ/cm$^{2}$ for the Si L$_{2,3}$ edge, respectively. 
Considering the reflectivity and absorbance coefficients of the overall sample stack at 1.55~eV \cite{refractiveindex} and limiting the absorption up to the first 100~nm of the Si substrate, the fraction of the total absorbed intensity of the incoming optical pump pulse released in the Ni layer is 95$\%$. 
The remaining energy is absorbed by the Ag capping layer (1$\%$) and the Si substrate (4$\%$). 
\hfill \\ \indent
The RMOKE analysis of the FEL light polarization angle $\theta$ as a function of the pump-probe time delay ($t-t_0$) is carried out with a Wollaston-like polarimeter, that collects the reflected FEL pulses \cite{Caretta21}. 
The polarimeter decomposes the polarization of the beam in two orthogonal components with I$_1$ and I$_2$ intensities. 
The polarization angle of the reflected beam with respect to the scattering plane is then approximated as \cite{donkerphd}
\begin{equation}
\theta= \dfrac{1}{2}\dfrac{I_1-I_2}{I_1+I_2}\,.
\end{equation}
\\
Element sensitivity of the Ni and of the Si layer is achieved by resonantly tuning the FEL radiation at the Ni M$_{2,3}$ edge \cite{Koide91} and the Si L$_{2,3}$ edge \cite{Antoniak12}. The applied magnetic field \textbf{B}, whose direction is parallel to the k-vector of the incoming FEL radiation, orients the magnetization of Ni along the line of intersection between the sample surface and the plane of incidence.
\begin{figure*}[ht!]
\captionsetup{justification=centerlast}
\centering
\includegraphics[width=0.95\textwidth]{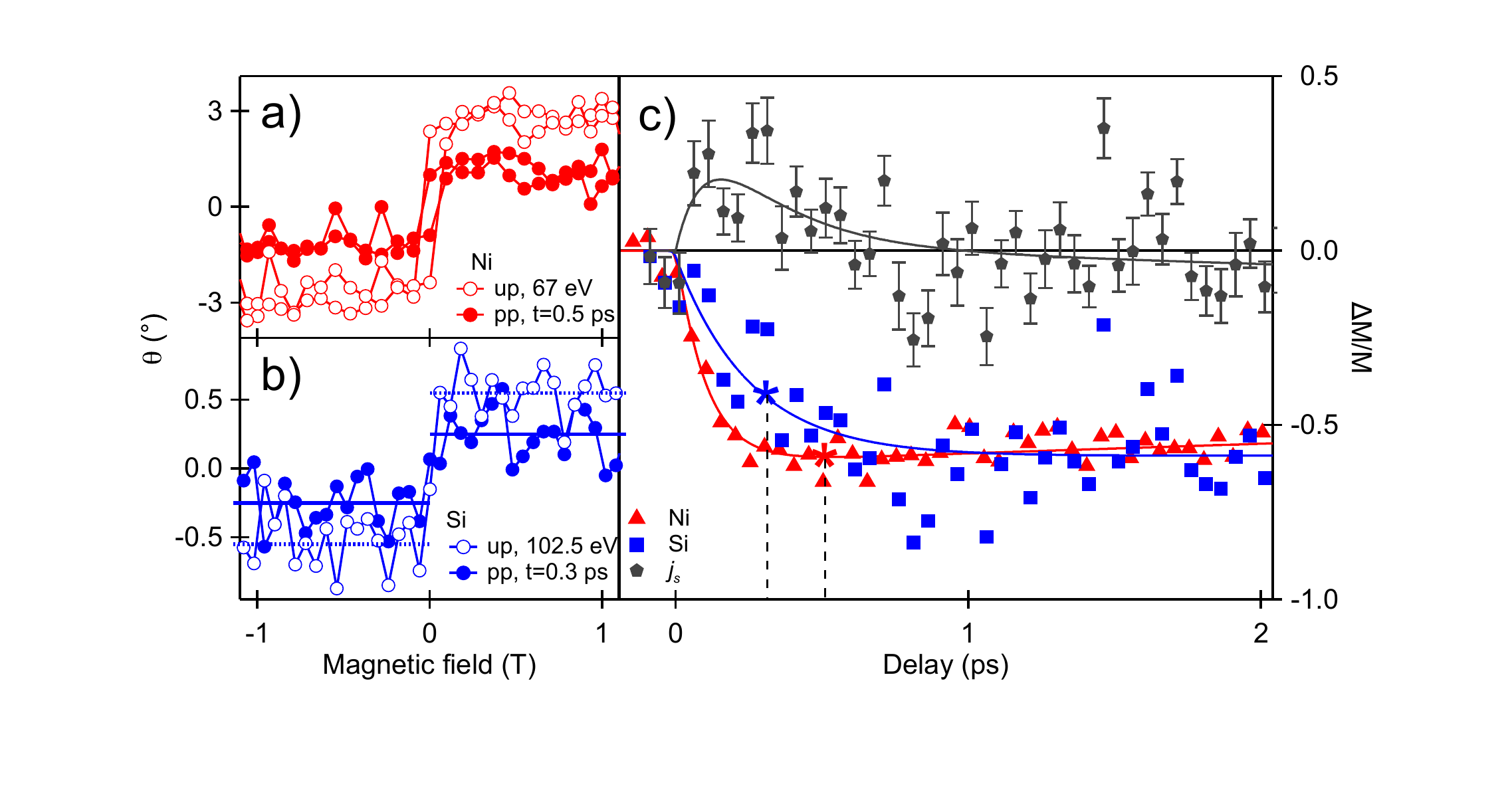}
\caption{\label{fig2}RMOKE magnetic hysteresis in degrees of Kerr rotation at the Ni M$_{2,3}$ edge (panel a) and at the Si L$_{2,3}$ edge (panel b).  
The empty and filled circles curves represent the unpumped (label $up$) and pumped (label $pp$) hysteresis measured at a delay time of 0.5 ps (for the Ni edge) and 0.3~ps (for the Si edge) (colored star markers in panel c). 
Panel c) relative change of the site resolved magnetization M (Ni - red dots and Si - blue dots) as a function of the time delay measured in saturation with an applied magnetic field of 550 mT. 
The Ni demagnetization curve was rescaled to account for the different pump fluence applied. 
The solid lines represent the fitting results, from which we extract the two characteristic times for demagnetization ($\tau_m$ ) and recovery ($\tau_r$). 
The difference of the two magnetization dynamics, defined as ($\Delta$M/M)$_{j_s}$, is also shown (gray pentagons).}
\end{figure*}

\hfill \\ \indent
Finally, the transient relative change of the sample magnetization magnitude M is defined as
\begin{equation}
\frac{\Delta M}{M}(t) = \frac{\theta(t)^+ - \theta(t)^-}{\theta_{sat}^+ - \theta_{sat}^-}\,,
\end{equation}
where $t$ is the delay time between the probe and the pump arrival, $\theta(t)^{\pm}$ are the RMOKE signals and $\theta_{sat}^{\pm}$ the RMOKE unpumped saturation values at opposite magnetic fields.

\section{\label{sec:res}Results}
Fig.~\ref{fig2}(a) displays the Kerr rotation collected at the Ni M$_{2,3}$ edge (67~eV) as a function of the applied magnetic field taken before the pump arrival (empty circles) and 500 fs after the pump absorption (filled circles). 
The silicon Kerr rotation collected with the FEL photon energy resonantly tuned to the Si L$_{2,3}$ edge (102.5~eV) and taken before and at a 300~fs time delay after the pump arrival are reported in Fig.~\ref{fig2}(b). 

The Kerr rotation displays a hysteresis shape for both the Ni and the Si edges. 
The difference of the respective saturation values $\theta_{sat}^{\pm}$$\mathrm{_{Ni}}$ and $\theta_{sat}^{\pm}$$\mathrm{_{Si}}$ originates from the different magneto-optical constants at the two absorption edges. 
The measured coercitivity field for Ni is $\sim$50~mT that confirms the ferromagnetic state of the Ni layer. 
Noteworthy, the Si Kerr signal measured at negative time delay reveals a non-vanishing magnetization at equilibrium in the silicon which is then partially reduced after the arrival of the pump pulse. 
This experimental finding will be further discussed below. \\
The reduced values of $\theta_{sat}^{\pm}$$\mathrm{_{Ni}}$ and $\theta_{sat}^{\pm}$$\mathrm{_{Si}}$ after the pump arrival prove the photoinduced partial demagnetization of both layers. 
It is interesting to note that the shapes of the demagnetized hysteresis are preserved indicating no detectable damaging of the sample caused by the pump pulse and the subsequent thermal heating. 
\hfill \\ \indent
Fig.~\ref{fig2}(c) displays the characteristic dynamics of the ultrafast relative change of the layer magnetization $\Delta$M/M measured at fixed FEL photon energies resonantly tuned to the Ni (red marks) and the Si (blue marks) edges with a saturating external field of 550~mT. 
The relative difference between the two dynamics is also reported (grey pentagons). 
In order to retrieve the characteristic lifetimes of the ultrafast magnetodynamics, the demagnetization curves were fitted using a decay-recovery double-exponential function
\begin{equation}
f(t)= \dfrac{\Delta M}{M} \Theta(t) \Big(1 -  e^{-t/\tau_m} \Big) e^{-t/\tau_r}\,,
\end{equation}
where $ \Theta(t)$ is the Heaviside step function, $\tau_m$ and $\tau_r$ are the demagnetization and the recovery times, respectively. Results of the fitting are shown in Fig.~\ref{fig2}(c) and summarized in Table~\ref{tab1} together with typical demagnetization and recovery times of Ni found in earlier experiments \cite{eschenlohr13, rudolf12}. 

\begin{table}[ht!]
\captionsetup{justification=centerlast}
\begin {center}
\begin{tabular}{c D{,}{\pm}{-1} D{,}{\pm}{-1}}\toprule
& \multicolumn{1}{c}{\textit{$\tau_m$ (fs)}} & \multicolumn{1}{c}{\textit{$\tau_r$ (ps)}}\\ \midrule
Ni & \textbf{100} \ , \ \textbf{12} & \textbf{20.3} \ , \ \textbf{5.6} \\
    & 140 \ , \ 10 ^{\text{\cite{eschenlohr13}}} & \multicolumn{1}{c}{\ \, \----} \\
    & 208 \ , \ 33 ^{\text{\cite{rudolf12}}} & 22 \ , \ 17 ^{\text{\cite{rudolf12}}} \\ \midrule
Si & \textbf{255} \ , \ \textbf{86} & \multicolumn{1}{c}{\ \, \textbf{$>$ 100}} \\ \bottomrule
\end{tabular}
\caption{\label{tab1} The fitting results for the demagnetization $\tau_m$ and recovery $\tau_r$ times obtained by the double-exponential function on Ni and on Si are reported. The results for Ni are compared with literature.}
\end {center}
\end{table}
\begin{figure}
\captionsetup{justification=centerlast}
\centering
\includegraphics[width=0.95\columnwidth]{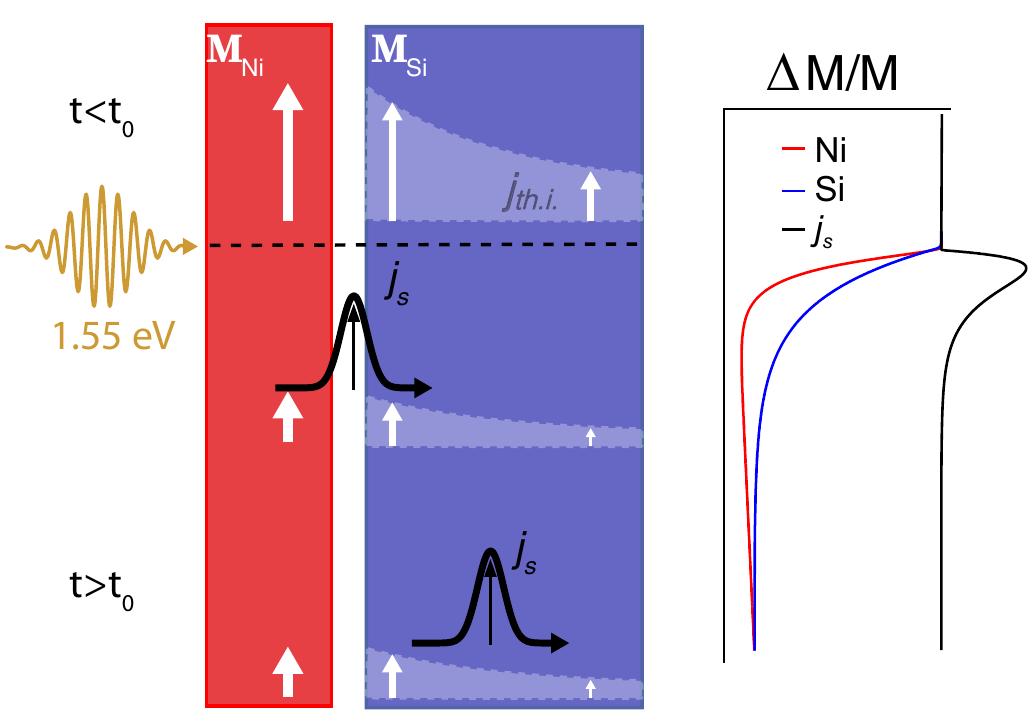}
\caption{\label{fig3}Sketch of the phenomena observed at the Ni/Si interface at timescales close to the IR pulse excitation. 
Top left, negative time delay $t<t_0$. 
In equilibrium conditions, the Ni layer (red) has a net magnetic moment \textbf{M}$\mathrm{_{Ni}}$ (white thick arrow). 
Due to the thermal excitation of the electrons in nickel, a spin polarized thermionic current $j_{th.i.}$ is established from the Ni to the Si layer. 
Accordingly, an exponentially decreasing magnetization profile \textbf{M}$\mathrm{_{Si}}$ (white thin arrows, lightened area) is established in silicon (blue). 
Lower left, IR pulse exciation and later times ($t>t_0$). 
Right after the arrival of the 1.55~eV pulse (golden wave at $t_0$), the magnetization of the Ni layer is quenched. 
The quenching of the magnetization in nickel triggers two phenomena. 
First, the reduction of the spin polarized thermionic current from the Ni to the Si layer, that finally leads to the reduction of the static magnetization of the Si layer. 
Second, the injection of a superdiffusive current ($j_s$, black pulse) into silicon, that propagates through the interface at later times. 
The two phenomena compete and cause the slow down of the demagnetization rate in silicon, as depicted in the right panel. 
Right panel; 
time dependent relative changes of the magnetization of the Ni layer (red), the Si layer (blue) and the contribution of the superdiffusive current $j_s$ (black). 
}
\end{figure}
\section{\label{sec:dis}Discussion}
The diagram shown in Fig.~\ref{fig3} summarizes our experimental findings and sketches the undergoing phenomena at the interface. 
The left panel shows the sample magnetization in the Ni film (\textbf{M}$\mathrm{_{Ni}}$) and in the Si substrate (\textbf{M}$\mathrm{_{Si}}$) before and after the pump arrival. 
The right panel represents the relative change of the Ni and Si magnetizations $\Delta$M/M and their relative difference as a function of the time delay.  
The initial state consists of a magnetized Ni film and a magnetized Si substrate as revealed by the measured RMOKE hysteresis reported in Fig.~\ref{fig2}(a) and (b). 
Unlike the Ni film case, the origin of the static magnetized state of the Si substrate is not trivial and it deserves a thorough explanation.
\hfill \\ \indent
\textbf{M}$\mathrm{_{Si}}$ originates from a proximity effect of the spin polarized electron distribution in the Ni film into the semiconductor. 
In fact, the magnetic Ni film acts as a reservoir of spin-polarized thermal electrons \cite{dankert13}. 
These low-energy electrons diffuse at equilibrium in the metallic layer and, due to the different exchange interaction with the spin-polarized electron background, they will be scattered differently depending on their spin. 
As a result, thermal electrons that impinge the metal/semiconductor interface are also spin-polarized and if their energy is above the Schottky barrier they can diffuse as a thermionic current ($j_{th.i.}$) inside the Si substrate \cite{dankert13}. 
This proximity effect is also called the spin-polarized thermionic effect at the metal-semiconductor interface \cite{dankert13}. 
Since the thermionic current decreases exponentially as a function of the distance from the metal/semiconductor interface, we also infer that the magnitude of the resulting magnetization has a depth dependence (shaded white profile in Fig.~\ref{fig3}, left panel). 
Accordingly, the proximal layer of the Si substrate retains a net collinear depth dependent magnetization because of the spin-polarized thermal electrons populating the semiconductor conduction band in the proximity of the Ni/Si interface. 
However, depth dependent investigation of the magnetization in Si is out of the scope of the present work.

\hfill \\ \indent
Now we turn to the discussion of the magnetodynamics observed at the Ni and Si edges.
The optical absorption of an ultrafast pulse by the Ni film is accompanied by a sudden increase of the electron temperature and a consequent reduction of the spin-polarization of the exchange-split Ni bands. 
In turn, while the Ni film demagnetizes in $\sim$100 fs, the consequent reduction of the spin-polarization of the Ni bands reflects in a reduction of the spin-polarized thermionic current $j_{th.i.}$ in Si. 
Accordingly, \textbf{M}$\mathrm{_{Si}}$ diminishes as revealed by the magnetodynamics at the Si edge. 
The most striking feature of Fig.~\ref{fig2}(c) is that within the first $\sim$1~ps the Ni and Si ($\Delta$M/M)(t) exhibit different dynamical responses to the laser excitation. 
Specifically, $\textbf{M}\mathrm{_{Ni}}$ reacts faster then $\textbf{M}\mathrm{_{Si}}$ and the resulting demagnetization rate of change in Ni is 2.5 times as fast as its Si counterpart. 
The difference ($\Delta$M/M)$\mathrm{_{Si}}$-($\Delta$M/M)$\mathrm{_{Ni}}$, that we indicate as ($\Delta$M/M)$_{j_s}$, is plotted in Fig.~\ref{fig2}(c) (grey pentagons) and it represents the transient spin current. 
In fact, as previously mentioned, since in thermal equilibrium the Ni spin minority-majority unbalance is at the origin of the spin-polarized thermionic effect in Si, the magnetodynamics measured at the two distinct absorption edges 
are expected to be identical and the difference between the two dynamics to be zero. 
However, as the Ni layer absorbs the femtosecond IR pump pulse, a spin-polarized current pulse $j_s$ is set and superdiffuses into the Si substrate through the interface. 
The propagation of $j_s$ is illustrated in the left panel of Fig.~\ref{fig3}. 
The spin polarization of $j_s$ reflects that of the spin majority in Ni. 
Consistently, as $j_s$ propagates into the Si substrate, it contributes as a transient increase of the Si magnetization, $\textbf{M}\mathrm{_{Si}}$. 
The combined effect between the spin polarized $j_s$ and $j_{th.i.}$ currents on the total transient magnetization of the Si substrate results in a longer demagnetization time $\tau_{m}$ of ($\Delta$M/M)$\mathrm{_{Si}}$ with respect to the Ni layer case. 
Consequently, based on this scenario it is justified to assume the difference between the Ni M$_{2,3}$ and Si L$_{2,3}$ magnetodynamics ($\Delta$M/M)$_{j_s}$ as an experimental evidence of the onset and propagation of a superdiffusive spin current across the Ni/Si interface. 
At a longer time scale, following the propagation of the spin current pulse inside the Si substrate, the observed ${j_s}$ contribution becomes irrelevant.

Finally, having established the presence of a spin current injected in the Si substrate across a Ni/Si interface, further quantitative information could also be extracted. 
($\Delta$M/M)$_{j_s}$ displays a maximum at $\sim$150~fs after the pump arrival, followed by an exponential decay time $\tau= 248~\pm~128$ fs, as obtained by an exponential fitting of the trace (see also the grey guide for the eye in Fig.~\ref{fig2}(c)). 
Considering the spin current pulse decay time $\tau=\Gamma/v$,  where $\Gamma$ is the optical attenuation length at the Si $ L_{2,3}$ edge, 
which in the present case is $\sim$55~nm \cite{Henke93}, the calculated velocity \textit{v} of the spin pulse propagating in the Si substrate results to be $\sim$0.2~nm/fs. 
This  experimental finding matches the theoretical predictions made on an ideal Ni/Si system (see Ref.~\cite{battiato16}).
\\
\section{\label{sec:conc}Conclusions}
In summary we experimentally observed magnetodynamics at the Ni M$_{2,3}$ and Si L$_{2,3}$ edges in a Ni/Si interface by making use of FEL-based element sensitive Kerr effect. 
The results indicate the onset of transient superdiffusive spin current triggered by an optical pump pulse from the ferromagnetic Ni layer into the Si substrate. 
The analysis of the magnetodynamics at the Ni  and Si edges reveals the slower demagnetization lifetime in Si when compared to Ni, which can be ascribed to the concomitant mechanisms of the thermionic proximity effect at the Ni/Si interface and the onset of a transient spin current triggered by an optical pump pulse.
Finally, the measured spin pulse propagation velocity into Si substrate reported here ($\sim$0.2~nm/fs) is benchmarking the theoretical values reported in literature \cite{battiato16,battiato10} giving further support to these models.

\section{\label{sec:ackn}Acknowledgments}
P.M and M.J. acknowledge financial funding though the project EUROFEL -- ROADMAP ESFRI.
%
%

\section{\label{sec:supple}APPENDIX}
\subsection{\label{sec:suppleA}Sample synthesis and characterization}
The sample was prepared and characterized under ultra-high vacuum conditions at the end-station of the VUV-Photoemission beamline (Elettra Sincrotrone Trieste). 
The Si substrate was cut from a p-doped Si wafer (B dopant, 0.05~$\Omega$cm resistivity) with (111) surface termination. 
After prolonged thermal annealing at 700~K, the substrate was flash-annealed to 1520~K until a sharp 7$\times$7 surface reconstruction appears in the low-energy electron diffraction (LEED) pattern. 
The topmost photoemission spectrum in Fig.~\ref{s2} (black line) refers to this preparation step. 
The surface was held at a temperature of 1050~K and exposed to 100~L of NH$_3$ to produce an ultra-thin layer of crystalline Si$_3$N$_4$(0001). 
This reaction is known to be self-limiting at the thickness corresponding to two bilayers ($<$~1~nm), which display an 8$\times$8 surface reconstruction in the LEED pattern \cite{Flammini15}. 
The formation of the Si$_3$N$_4$ layer is accompanied by the presence of the N$1s$ level and the shifted components in the Si$2p$ and Si$2s$ levels (red spectrum in Fig.~\ref{s2}). 
Ni was grown on the Si$_3$N$_4$(0001) surface kept at liquid nitrogen temperature, to favor the formation of a continuous film. 
As shown in Fig. \ref{s1} (green spectrum), the Ni film thickness of 7~nm is sufficient to suppress almost completely the signal from the Si$_3$N$_4$(0001). 
The Ni film displays no LEED pattern. 
Finally, an Ag film of 2~nm was deposited on Ni, again using liquid nitrogen temperature to obtain a uniform coverage (blue spectrum of Fig. \ref{s1}). 
This was particularly important to prevent the oxidation of Ni, as the sample was removed from the growth chamber for the magnetic measurements.

\subsection{\label{sec:suppleB}TEM characterization of the sample}
\begin{figure}
\captionsetup{justification=centerlast}
\centering
\includegraphics[width=0.95\columnwidth]{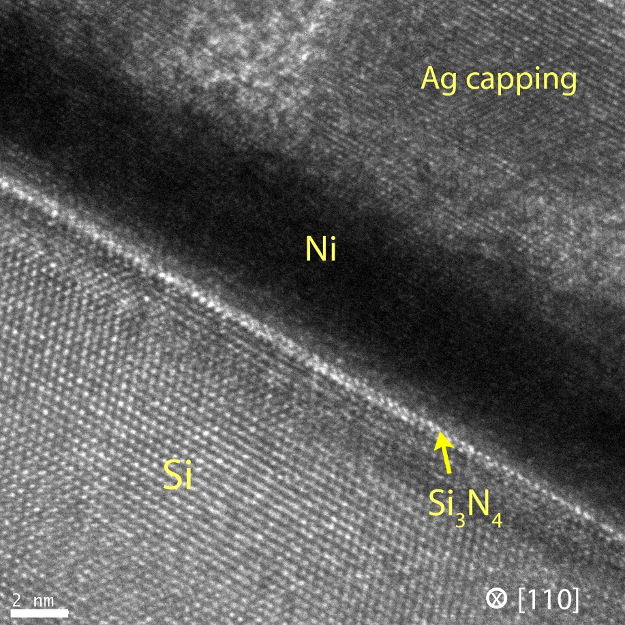}
\caption{\label{s2}Cross-sectional HRTEM image of Ni/ Si$_3$N$_4$ /Si (111) nanostructure.}
\end{figure}

\begin{figure}
\captionsetup{justification=centerlast}
\centering
\includegraphics[width=0.95\columnwidth]{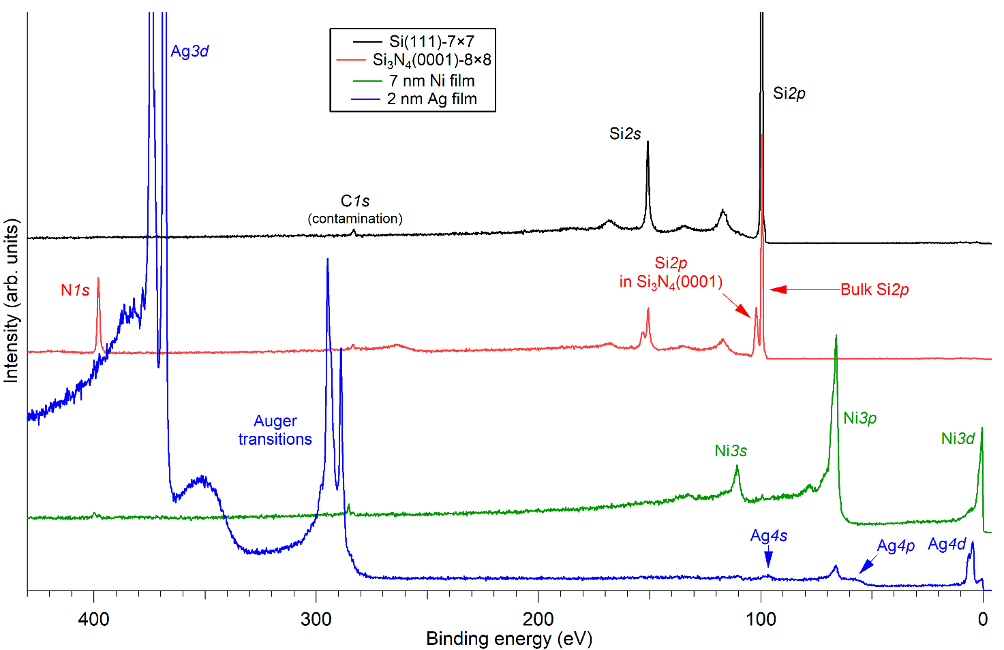}
\caption{\label{s1}Photoemission spectra taken at the various stages of the sample growth.}
\end{figure}

Details on the film nanostructure were provided by high-resolution transmission electron microscopy (HRTEM) using a JEOL 2010 UHR field emission gun microscope operated at 200~kV with a measured spherical aberration coefficient $C$s of 0.47$\pm$0.01. 
A representative bright-field HRTEM image of the Ni/ Si$_3$N$_4$/Si(111) cross-section is shown in Fig.~\ref{s2}. 
The Si$_3$N$_4$ layer is about 0.7~nm-thick and homogeneously extends on top of the bare Si (111) substrate with atomically flat and sharp interfaces.


%
%


%

\end{document}